\documentclass[fleqn,12pt,twoside]{article}
\usepackage[headings]{espcrc1}

\readRCS
$Id: espcrc1.tex,v 1.2 2004/02/24 11:22:11 spepping Exp $
\usepackage{graphicx}
\usepackage{epsfig}
\usepackage[figuresright]{rotating}

 \newcommand{\beq}{\begin{equation}}
 \newcommand{\eeq}{\end{equation}}
 \newcommand{\bea}{\begin{eqnarray}}
 \newcommand{\eea}{\end{eqnarray}}
  \newcommand{\bit}{\begin{itemize}}
 \newcommand{\eit}{\end{itemize}}

\newcommand{\AmS}{{\protect\the\textfont2
  A\kern-.1667em\lower.5ex\hbox{M}\kern-.125emS}}

\title{TOWARDS A NEW STATE OF MATTER}

\author{Luciano Maiani \address[ULS]{Universit\`{a} di Roma La Sapienza and INFN, Roma, Italy.}%
        \thanks{Work supported in part by Ministero Istruzione, Universit\`{a} e Ricerca and by Istituto Nazionale di Fisica Nucleare, Italy }.}

\runtitle{Towards a new state of matter}
\runauthor{L. Maiani}

\begin{document}

\maketitle

\begin{abstract}
Progress on QGP
\end{abstract}

\section{INTRODUCTION}

The issue of hadronic matter at high temperature was raised in the study of the Early Universe, at least this is where I met it first.  

Quoting from the book of S. Weinberg  \cite{WEIN72}:  {\it  if we look back ...when temperature was above $10^{12}\;^0K$ (100 MeV), we encounter theoretical problems of a difficulty beyond the range of modern statistical mechanics..}. There were, at the time, {\it two extremely different simple models...the hope is that one or the other may come close enough to reality to lead to useful insights about the very early Universe}.

The first model was the so-called {\it Bootstrap} originated in the sixties and evolving in the Dual Models, in the early seventies: all hadrons are composite of one another (a concise summary of the early Veneziano and Dual Models and the role of an infinitely rising spectrum is found in~\cite{Green}). The second model, just taking over in these years in the form of the present Standard Model, was the {\it Elementary Particles} model. Matter and radiation are made of elementary constituents: photons, leptons, quarks, gluons, which behave as free particles at high energy (asymptotic freedom). As we know today, both models may be right, although in different energy and temperature ranges.

The bootstrap hadron picture applies in the low energy, strongly interacting, regime, giving rise to 
an almost exponentially rising hadronic level spectrum and to a limiting temperature that can be estimated from the hadron spectrum, the Hagedorn temperature, $T_H=$~170-180 MeV. Just before reaching the limiting temperature, hadrons {\it melt} in a deconfined state made by the elementary constituents (the Quark Gluon Plasma, QGP in brief). 

There are still many open questions in this field:
\begin{itemize}
\item
is deconfinement a true phase-transition or simply a smooth cross-over?
\item
in the latter case, is there a tri-critical point where hadronic and  QGP phases coexist?
\item
is chiral symmetry restored in the deconfined phase? 
\item
after the transition, do we find an almost  free gas, or rather is there an intermediate phase of strongly interacting Quark Gluon Plasma (sQGP)?
\end{itemize}

\begin{figure}[ht]
\begin{center}
\epsfig{
height=7truecm, width=7.5truecm,
        figure=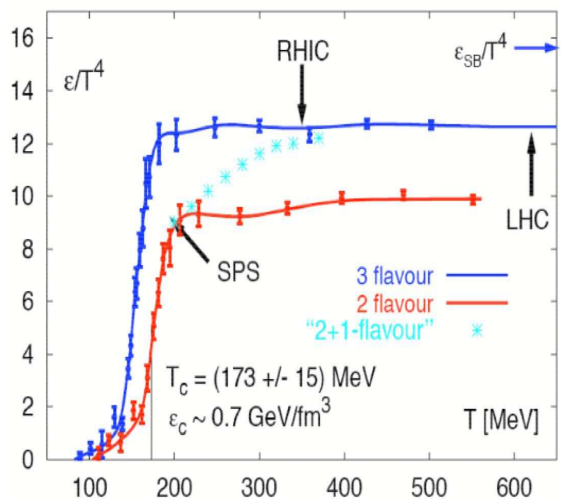}
\caption{\label{fig PPP} \footnotesize Energy density, $\epsilon$, vs. Temperature from lattice QCD calculations~\cite{latKar}. For convenience the ratio $\epsilon/T^4$ is plotted.}
\end{center}
\end{figure}

An important theoretical guidance is provided by Lattice QCD calculations, which indeed show a rapid rise in the energy density of hadronic matter around $T\simeq$170~MeV and energy density $\epsilon \simeq$~2GeV/fm$^3$~\cite{latKar}, at baryon chemical potential, $\mu_B=0$ (see Fig.~\ref{fig PPP}). The last years have seen many attempts to explore theoretically the $\mu_B\neq 0$ region, with a variety of methods that will be reviewed in this Conference and that witness the vivacity and interest of this field.
%

Experiments with Heavy Ion Collisions, performed over the last decade at the AGS (Brookhaven), SPS (CERN) and RHIC (Brookhaven) have addressed these issues with increasing depth and precision, as this Conference will show, and provided us with an exciting mix of expected results and big surprises. 

\begin{figure}[htb]
\begin{minipage}[t]{80mm}
\epsfig{
height=7.6truecm, width=8.6truecm,
       figure=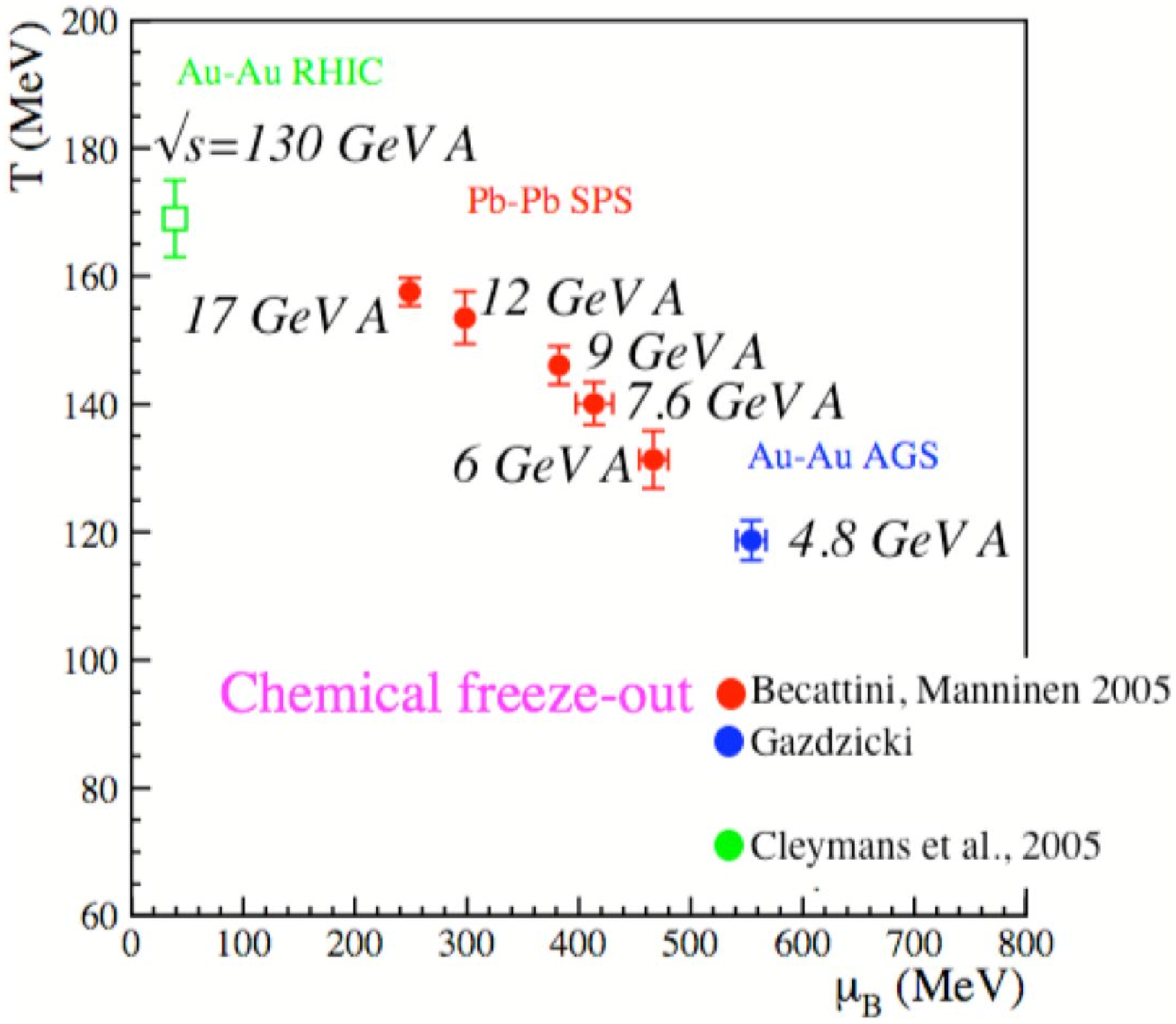}
\caption{\label{fig MMM} \footnotesize Temperatures vs. baryon chemical potential, $\mu_B$, for particles at freeze-out in present Heavy Ion facilities }
\end{minipage}
\hspace{\fill}
\begin{minipage}[t]{75mm}
\epsfig{
height=7truecm, width=8truecm,
        figure=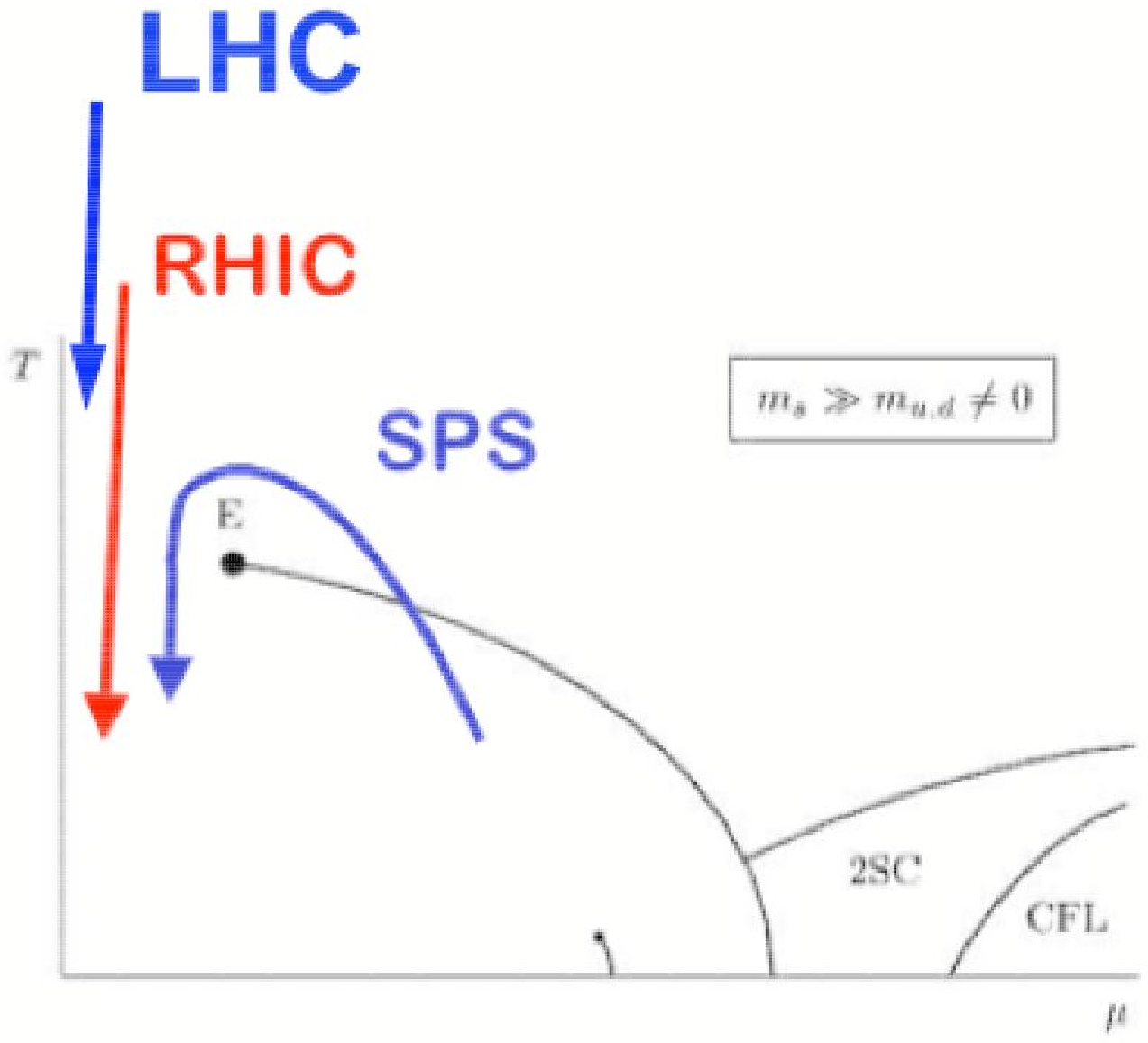}
\caption{\label{fig TH} \footnotesize Phase diagram of hadronic matter in the T-$\mu_B$ plane: pictorial view of the present theoretical expectations.}
\end{minipage}
\end{figure}

Fig.~\ref{fig MMM} shows the location of each series of experiments in the ($T-\mu_B$) plane. The points correspond to the hadrons at the chemical freeze-out point, the final stage of the fireball created in the collision, when hadrons cease to interact and fly undisturbed towards the detectors. 
Fig.~\ref{fig TH} shows a pictorial view of the phase diagram corresponding mostly to our theoretical guesswork, with superimposed the location of the regions within reach of the SPS, RHIC and the LHC. 

Remarkably, the working points of the accelerators go across the region where the transition should occur, according to the theoretical investigations. With increasing energy, the freeze-out points head towards vanishing $\mu_B$, the domain of the Early Universe, which will be explored exhaustively with the LHC. The study of QGP is thus a fertile ground for theory-experiment cross talk, which makes it particularly fascinating and rather unusual in today's High Energy trade.

I was asked by the Organizers to comment on the SPS results, five years after their presentation at CERN. There are at least  two other good reasons to do so. 

This year, the results of the second generation experiment following the first SPS campaign, NA60, have been presented.  It is  illuminating to confront the new data with the older ones, to get a better picture. 

The second reason is that you will hear exhaustively at this Conference about RHIC results, which dominate the scene today, so I do not have to feel guilty if, in the search of a new state of matter, I shall concentrate on the onset of the transition, the lower energy region where the SPS seems to be luckily sitting.

\section{THE SPS CAMPAIGN AT CERN}

The Heavy-Ion facility  at the CERN SPS was  realized as a collaboration between CERN 
and the laboratories reported in Table~\ref{SPSconstr}. It was a good example of an international collaboration to build a new facility, a model for the LHC and for future High Energy facilities. The SPS provided ion beams to several experiments  in the North Area.

\begin{table}[htb]
\caption{Participanting institutions in the construction of the SPS Heavy Ion facility}
\label{SPSconstr}
\newcommand{\m}{\hphantom{$-$}}
\newcommand{\cc}[1]{\multicolumn{1}{c}{#1}}
\renewcommand{\tabcolsep}{2pc} 
\renewcommand{\arraystretch}{1.2} 
\begin{tabular}{@{}lll}
\hline
GANIL               & Caen & FRANCE  \\
INFN          	& Legnaro	& ITALY \\
Universit\`{a}			& Torino	&ITALY \\
GSI			& Darmstadt	& GERMANY \\
Institute of Applied Physics	& Frankfurt & GERMANY\\
Variable Energy Cyclotron (VECC)&  Calcutta	& INDIA\\	
Baba Atomic Research Centre  (BARC) & Mumbay	 &INDIA \\
Tata Institute (TIFR)	& Mumbay	& INDIA \\
Academy of Sciences	& Prague	& CZECH REP \\
\hline
\end{tabular}\\[2pt]
In-cash contributions from: SWEDEN and SWITZERLAND
\end{table}

The main campaign of data taking at the SPS involved four detectors operated by large international collaborations: NA50, dedicated to the observation of $\mu$ pairs at high energy, in particular in the J/$\psi$ region, NA57 (previously WA97) and NA49, focussed on strange particle production, and NA45 (CERES) on low energy electron pairs. The campaign ended in 1999, after which only one new experiment, NA60, was put on the floor by a continuation of the NA50 Collaboration. This year the NA60 results, really of second generation quality,  have been presented, for the first time, at the EPS Conference in Barcelona and now here at Quark Matter.

 The results of data collection at the SPS were summarized in February 2000: {\it .. data provide evidence for colour  deconfinement in the early collision stage and for a collective explosion of the collision fireball in its late stages. The new state of matter exhibits many of the characteristic features of the theoretically predicted Quark-Gluon Plasma} ... {\it The challenge now passes to the Relativistic Heavy Ion Collider at Brookhaven and later to CERN's Large Hadron Collider.}
 
\section{CRITICAL LENGTHS FOR J/$\psi$ ABSORPTION IN HEAVY NUCLEI}
\begin{figure}[htb]
\begin{minipage}[t]{80mm}
\epsfig{
height=5.3truecm, width=7.3truecm,
       figure=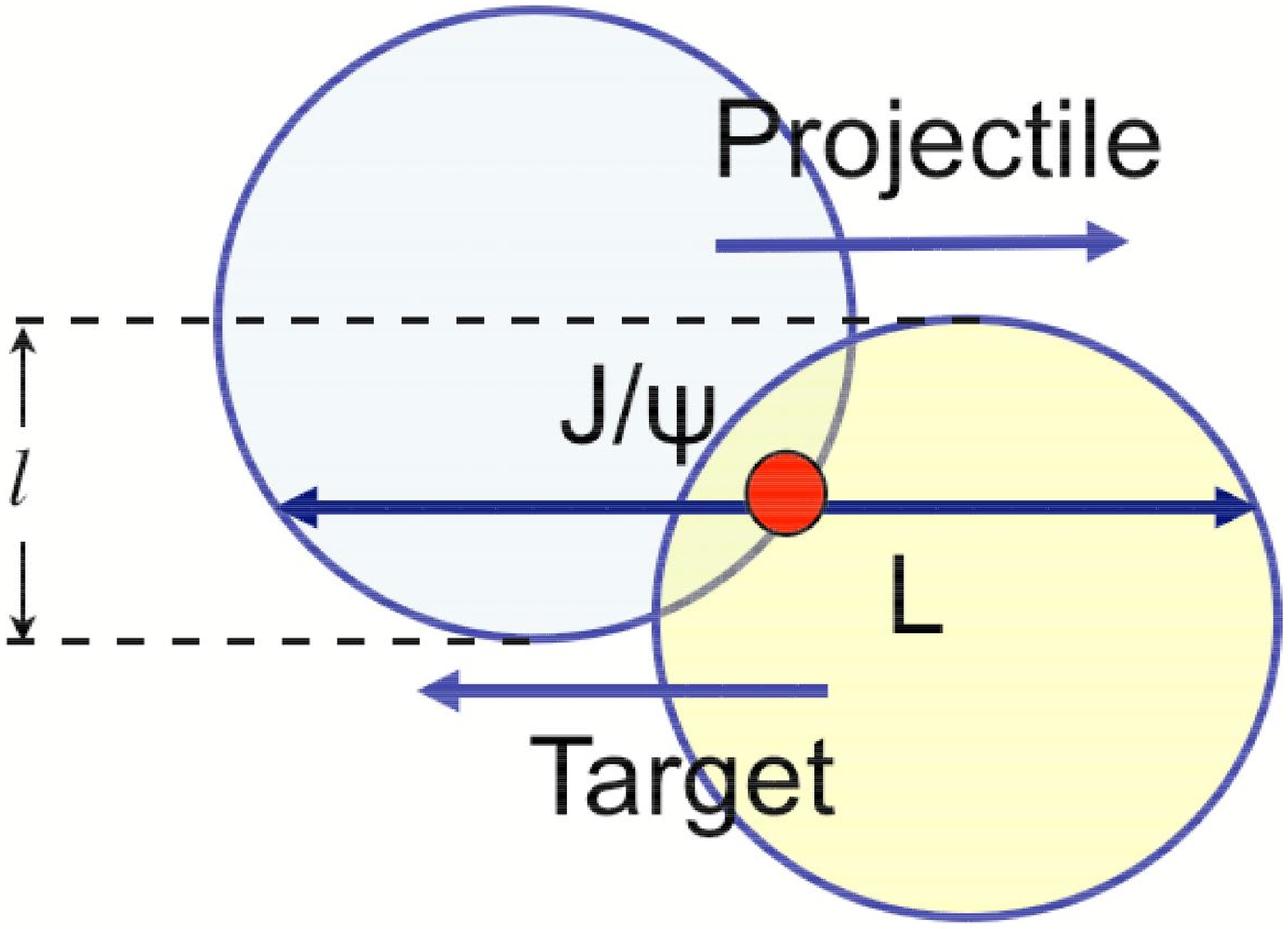}
\caption{\label{fig LLL} \footnotesize The two characteristic lengths in Heavy ion collisions. }
\end{minipage}
\hspace{\fill}
\begin{minipage}[t]{75mm}
\epsfig{
height=7.5truecm, width=8.4truecm,
        figure=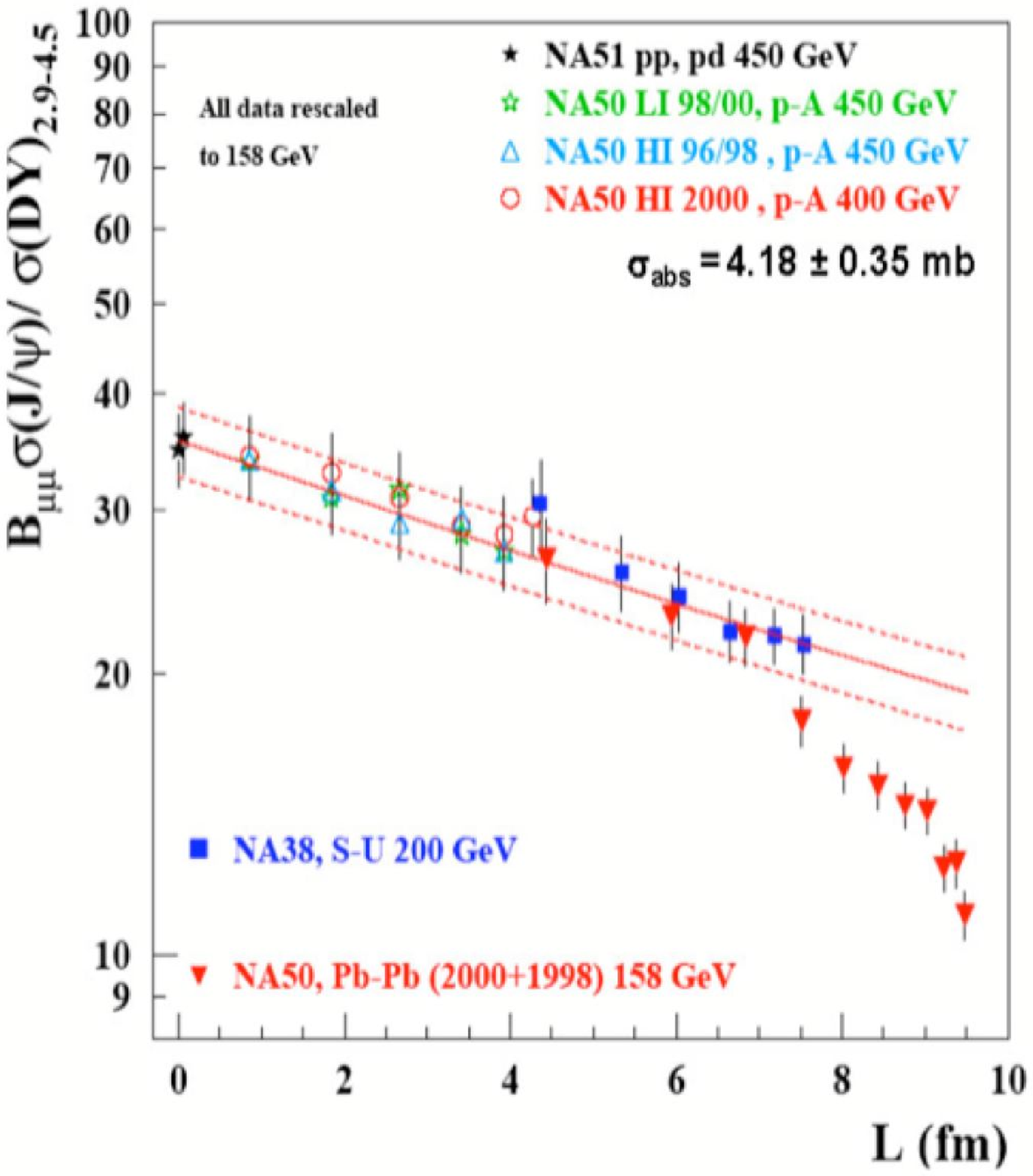}
\caption{\label{fig versusL} \footnotesize J/$\psi$ production rate normalized to Drell-Yan vs. $L$. Straight lines represent nuclear absorption. Data from NA38 and NA50, see~\cite{NA50Pb,NA60In}.}
\end{minipage}
\end{figure}

There are two characteristic lengths in the process:
\beq
A\;+\;A'\to J/\psi\;+\;\rm anything
\eeq
namely (see Fig.~\ref{fig LLL}):
\begin{itemize}
\item
the length of the nuclear matter column, $L$, that the J/$\psi$ has to traverse during the collision; $L$ determines the nuclear absorption coefficient, which is given by $\sigma_{\rm abs} \rho_{\rm nucl}$L and $\sigma_{\rm abs}$ is determined experimentally by studying J/$\psi$ production in $p+A$ reactions as function of A (in the c.o.m. the length is L/$\gamma$ but $\rho$ is increased by the same Lorentz factor so that the attenuation is boost independent);
\item
the second one, denoted by $\it l$, is the initial transverse size of the fireball; $\it l$ determines the absorption of the J/$\psi$ by the mid-rapidity hadrons (the {\it comoving particles}). 
\end{itemize}

The transverse size, $\it l$ is related to the impact parameter (for equal nuclei, $\it l =2R-b$, where $R$ is the nuclear radius) and also to the commonly used number of participant nucleons, $N_{part}$ (for small $\it l$, $N_{part}\simeq const.\cdot {\it l}^3$). 
For given nuclei, $L$ is also a function of $\it l$ that can be computed, in the approximation of sharp spheres or with Wood-Saxon distributions, with very similar results. 

In fact, $L$ is maximum for the largest centrality, $\it l=2R$. This determines the very peculiar turning down in Fig.~\ref{fig versusL}. This feature simply signals non-vanishing absorption by comovers: 
\beq
\frac{\partial}{\partial L}(\rm Rate)=\frac{\partial}{\partial {\it l}}(\rm Rate)\cdot \frac{1}{(\partial L/\partial \it l)};\;\;i.e. \;\; \frac{\partial}{\partial L}(\rm Rate)\to -\infty,\;\;if\;\;\frac{\partial}{\partial {\it l}}(\rm Rate)<0
\eeq

$\it l$ is a good candidate to be the critical parameter. The transverse length determines the linear size of the initial fireball and therefore its initial volume. It is most reasonable that collective phenomena like the ones we are addressing can take place {\it only} when the hadron matter involved reaches a certain minimum volume. In addition, the number of nucleons per unit surface which take part in the collision increases with $l$. Related to the energy density, the temperature of the fireball also increases: increasing centrality (i.e. $\it l$) we make a temperature scan which may cross the critical temperature.

\section{LIMITING TEMPERATURE AND J/$\psi$ DISSOCIATION}
The basic cross sections for: 
\beq
\pi/\rho+ J/\psi \to\;open\;charm
\eeq
are large (few to 10 mb) and strongly energy dependent, due to the proximity of several thresholds (see e.g.~\cite{brambilla}). Assuming a thermalized fireball, its opacity to the J/$\psi$ increases strongly with temperature: we could account for a large absorption with correspondingly large T (not a discontinuity, which is, however, more difficult to prove). 
\begin{figure}[ht]
\begin{center}
\epsfig{height=7truecm, width=7.5truecm,
        figure=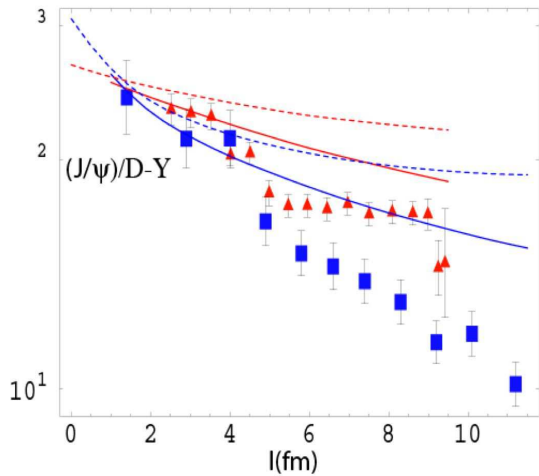}
\caption{\label{fig PBIN} \footnotesize $J/\psi$ production, normalized to Drell-Yan pairs, vs $l$ in Pb-Pb (boxes, NA50) and In-In (triangles, NA60) collisions. Dashed lines represent the expected values after nuclear absorption correction~\cite{NA60In}; solid lines include the hadronic absorption, computed assuming a limiting Hagedorn temperature: $T_{Hag}=180$~MeV~\cite{MPPR_NuclPhys} (upper = In, lower = Pb, for both dashed and solid lines).}
\end{center}
\end{figure}

A limiting temperature of hadronic matter around 170-180 MeV is suggested by independent theoretical arguments and is supported by the approximate exponential behavior of the hadron level spectrum and by observed limiting features in the hadronization process~\cite{Bec}. In turn, a limiting temperature implies, most likely, a limiting absorption. Thus, hadronic matter only cannot explain the observed opacity, if the latter turns out to exceed the limiting absorption itself~\cite{MPPR_NuclPhys}.

Thermalization of the initial fireball is an important issue, that deserves consideration ''per se", because of its implications on other aspects of the data, such as the elliptic flow of the outgoing particles. Data gathered at RHIC speak in favor of very early thermalization. This should apply to SPS as well, where hydrodynamic elliptic flow has been reported~\cite{ellfloSPS}. 

Fig. \ref{fig PBIN} shows a recent analysis of the NA50 (Pb-Pb)~\cite{NA50Pb} and NA60 (In-In)\cite{NA60In} data performed by our group along these lines~\cite{MPPR_NuclPhys}. Assuming a limiting temperature of $180$~MeV, we see that the nuclear+hadronic absorption by a hadron gas falls short from reproducing the observed drop beyond $\it l=$4-5 fm. 

\begin{figure}[htb]
\begin{minipage}[t]{80mm}
\epsfig{height=6.3truecm, width=7.5truecm,
       figure=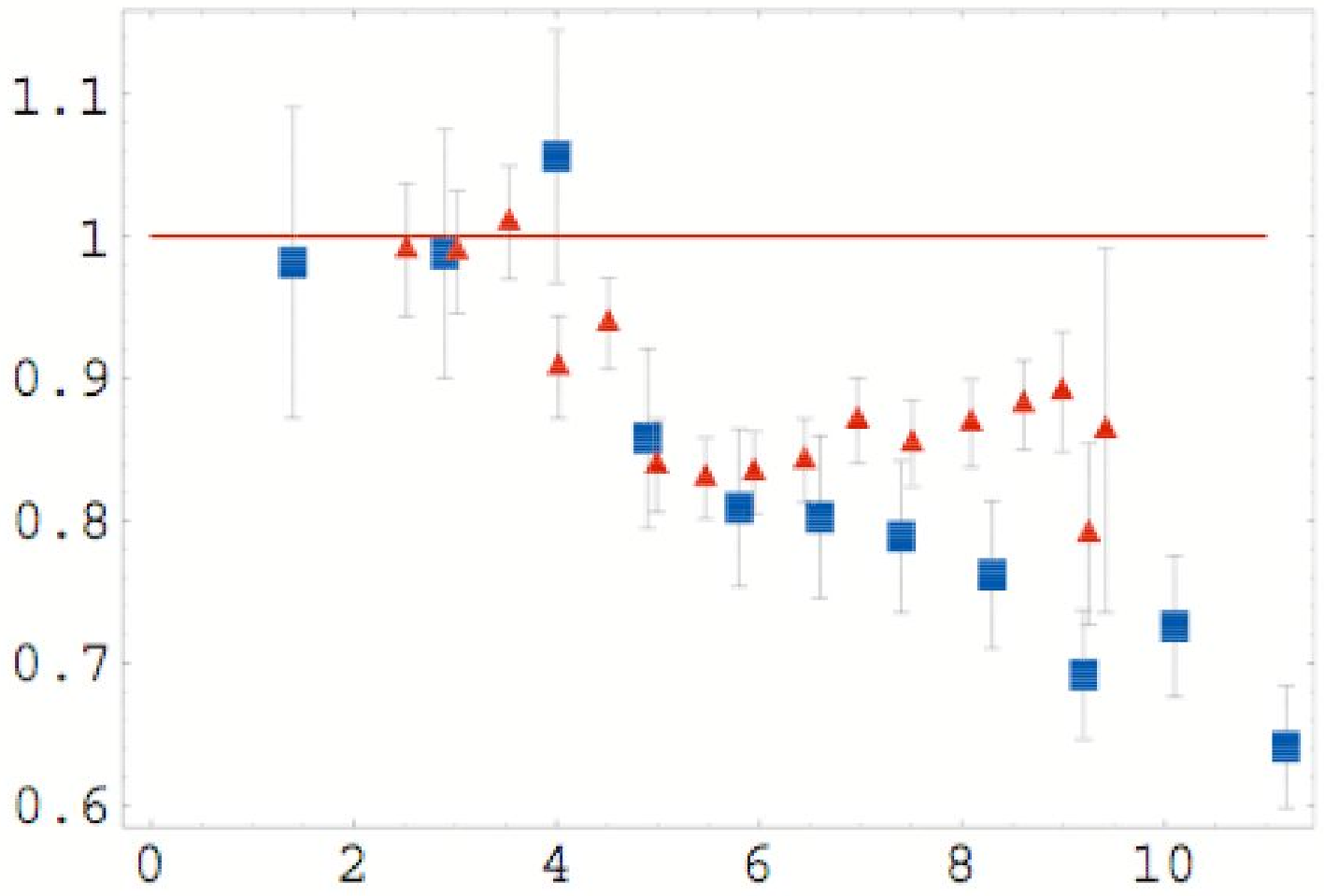}
\caption{\label{fig Z} \footnotesize {\it Observed/Expected} vs. $\it l$.}
\end{minipage}
\hspace{\fill}
\begin{minipage}[t]{75mm}
\epsfig{height=6.3truecm, width=7.5truecm,
        figure=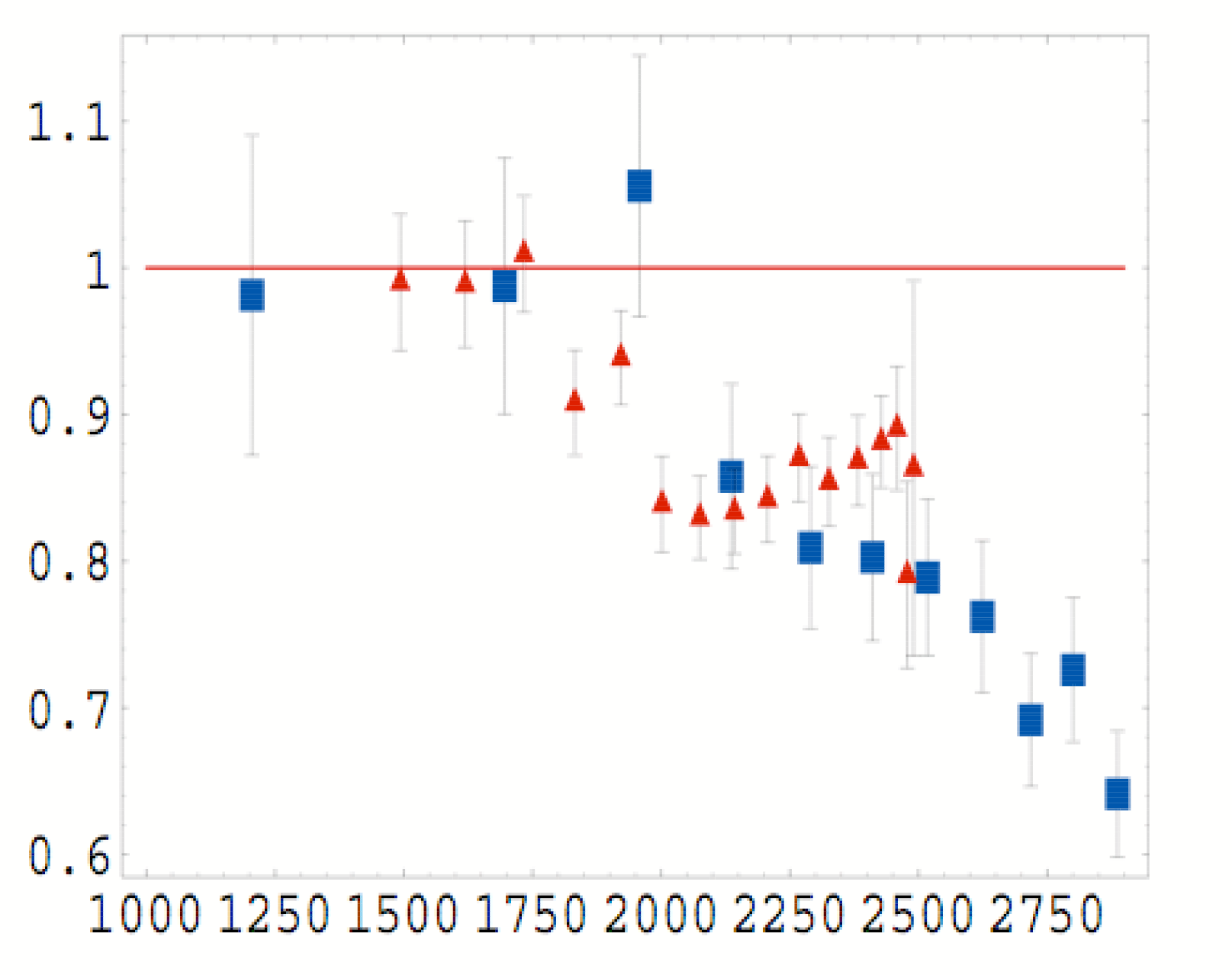}
\caption{\label{fig enden} \footnotesize {\it Observed/Expected} vs Energy density. We assume that $\it l=4$ corresponds to $\simeq 2$~GeV/fm$^3$ and assign the energy density to the other points using the Bjorken formula~\cite{Bj}.}
\end{minipage}
\end{figure}

The ratio {\it Observed/Expected} vs {\it l} (Fig.~\ref{fig Z} ) shows a clear discontinuity around $\it l=$4-5 fm. There is resonable agreement between Pb and In data.  If the driving consideration is the energy density produced by the collision, the step should occur later in $\it l$ for Indium, due to the lower baryon surface density at given $\it l$~\cite{Bj} (a similar conclusion is found in the percolation model~\cite{Satzetal}). 

In fact, the agreement between In and Pb data is slightly better if we plot the same ratio versus the energy density, computed by fixing $\epsilon=$2~GeV/fm$^3$ at $\it l$=4 fm and scaling $\epsilon$ to other values of $\it l$ with the baryon surface density, as implied by the Bjorken formula~\cite{Bj} (see Fig.~\ref{fig enden}). The scale of the energy densities agrees reasonably with those expected for the disappearence of the higher charmonium states, $\psi'$ and $\chi_c$~\cite{screening}, which are an important source of J/$\psi$~\cite{ioffe}. 

\section{STRANGENESS ENHANCEMENT AT THE SPS}

Primary strange particle densities in phase-space are parameterized according to:
\begin{equation}\label{gammas}
\rho_S(E,T)=(2J+1)(\gamma_S)^{n_S}\;e^{\frac{{\vec \mu}\cdot {\vec q}-E}{T}},
\end{equation}
where $n_S$ is the number of valence strange quarks and antiquarks in the particle, ${\vec q}$ are a set of conserved charges, ${\vec \mu}$ the corresponding chemical potentials and $\gamma_S$ is the strange particle suppression factor. But $\gamma_S$,.. what is it ?

The rationale usually given is that $\gamma_S$ summarizes the deviation from statistical equilibrium of strange particles, due to a ''lack of time" for equilibrating strangeness 
from the initial state which contains very few strange quarks. 

The standard picture, then, is that strange quarks should equilibrate better in the deconfined phase, because of the small strange quark current mass, and strange hadrons form more efficiently from quark recombination at freeze-out. Hence, the  Ò fudge factor Ó $\gamma_S$ approaching unity should be a good indicator of QGP formation. Indeed thermodynamical fits of particle abundances at RHIC find $\gamma_S\simeq 1$~\cite{BraunMunz}. 

In this framework, one may search for a correlation between J/$\psi$ suppression and the observed increase of strange particle production observed at the SPS. The previous discussion suggests to look for a correlation of $\gamma_S$ with centrality. Ratios of strange to non-strange particle production as function of centrality have been published recently by NA49~\cite{NA49_05} and show indeed a rapide increase for $N_{part}\simeq80$. We have followed the suggestion made in~\cite{MaiCatania}, to look for the behaviour of $\gamma_S$ as function of $\it l$ and correlate it to J/$\psi$ suppression.
 
\begin{figure}[ht]
\begin{center}
\epsfig{
height=8truecm, width=8.5truecm,
        figure=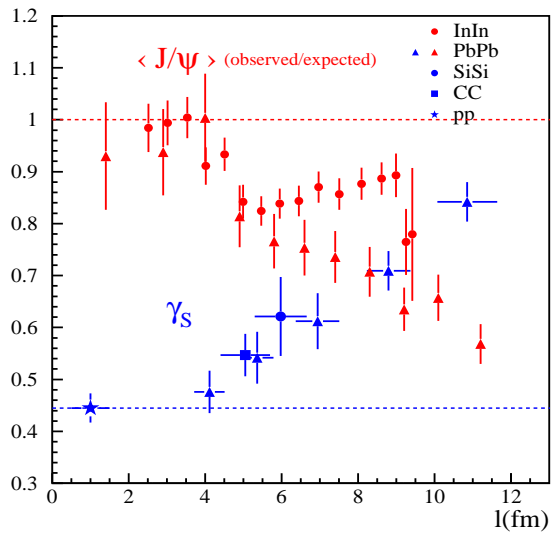}
\caption{\label{fig TTT} \footnotesize Strangeness undersaturation
factor, $\gamma_S$,  and $J/\psi$ anomalous suppression ratio, $R_{J/\psi}$, as functions of the transverse size of the interaction region in heavy ion collisions at $\sqrt{s}_{\rm NN}=17.2$~GeV (see text for definitions). $J/\psi$ points are from Pb-Pb (NA50) and In-In (NA60) collisions (triangles and circles, respectively, with vertical error bars only); values of $\gamma_S$ refer to Pb-Pb, Si-Si, C-C and p-p collisions (triangles, circle, box, star, with vertical and horizontal error bars), data from NA49 and NA57.}
\end{center}
\end{figure}

Our main result~\cite{Bec_al_05} is shown in Fig.~\ref{fig TTT}, where $\gamma_S$ is reported as function of $\it l$, together with the double ratio:
\begin{equation}
R_{J/\psi}=\frac{[Rate(J/\psi)/(D-Y)]_{\rm Observed}}
{[Rate(J/\psi)/(D-Y)]_{\rm Expected}}.
\label{errePsi}
\end{equation}
Indeed, $\gamma_S$  departs from its low value in proton-proton or light nuclei collisions to approach unity in the same centrality range where $R_{J/\psi}$ drops below unity. 

The correlation we observe in Fig.~\ref{fig TTT} makes it considerably stronger the case for the SPS being right at the onset of QGP formation. Further experimental investigations in this energy range are clearly called for, to elucidate the nature of the transition.  

\section{GLIMPSES OF STRONGLY INTERACTING QGP}

Main findings at RHIC can be summarized as follows~\cite{shurLNF}:
\begin{itemize}
\item
Particles are produced from matter which seems to be well equilibrated by the time it gets to chemical freeze-out: particle abundances follow the Boltzmann law and $\gamma_S\simeq 1$;
\item   
Very robust collective flows were found, indicating very strongly coupled Quark-Gluon Plasma (sQGP); 
\item
Strong quenching of large $p_T$  jets: they  are strongly absorbed by the fireball. The deposited energy seems to go into hydrodynamical motion (conical flow);
\end{itemize}

Early thermalization and low viscosity suggest that above the transition there is a strongly interacting quark-gluon {\it liquid},  rather than a weakly interacting QGP, perhaps coexisting with the most tightly bound hadrons ($\rho$, $\omega$, $\phi$, J/$\psi$). 

In a colored medium, color forces are screened and there is not a strong difference between color singlets and non-singlets - at least for heavy quarks. It is an intriguing speculation the possibility that there may be color non-singlet bound states as well~\cite{shurLNF}.
\begin{figure}[ht]
\begin{center}
\epsfig{
height=7.5truecm, width=7.5truecm,
        figure=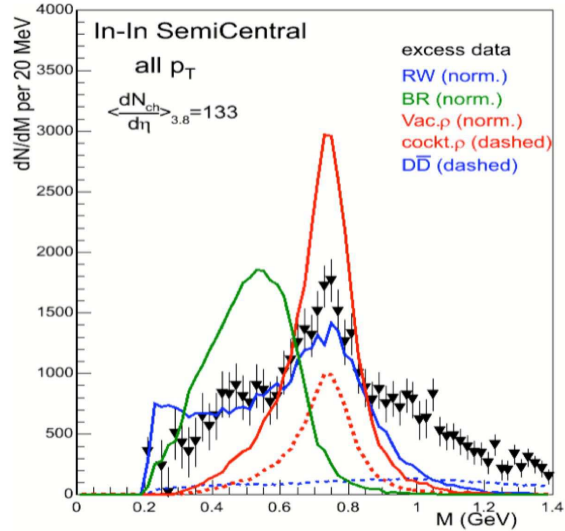}
\caption{\label{fig VVV} \footnotesize Dimuon spectrum after subtraction of the {\it cocktail} distribution of known resonaces except the $\rho$. Data from NA60 presented at this Conference~\cite{DamjaQM05}. Solid red line: the $\rho$ line shape expected in absence of effects from the medium.}
\end{center}
\end{figure}

One possibility to reconcile the sQGP with the precocious asymptotic freedom that we {\it do see} in deep inelastic scattering processes is the fact that temperature increases very slowly with the energy density of the medium, in fact like $\epsilon^{1/4}$ (unlike $Q^2$ in deep inelastic scattering) so that even at RHIC we happen to be still very near to the transition region. If this is indeed the case, experiments at the SPS can be useful, to shed light on the bound states that may populate the sQGP.

An intriguing example of effects of the medium is provided by the NA60 data on the distribution of low energy dimuons, which confirm and extend the previous NA45(CERES) results. Fig. \ref{fig VVV} shows the data after subtraction of all  signals expected from  known sources (the {\it cocktail}) but the $\rho$~\cite{DamjaQM05}. If we assume that the residuum is indeed the $\rho$ signal and compare it with the same signal {\it in vacuo}, we see a considerable broadening (in agreement with some models~\cite{RW}) but no shift in the resonance mass (in contrast with other models~\cite{BR}). 
 
 Definitely, we need to know more about this issue, {\it at both RHIC and the SPS}.

\section{CONCLUSIONS}

All indications are that deconfinement is seen at the SPS, in the collisions with largest centrality. Strangeness enhancement and J/$\psi$ suppression are correlated in centrality, the $\rho$ spectral function is modified w.r.t. normal vacuum.

If this is so, the SPS offers the unique possibility to study the onset of deconfinement and give significant contribution also to the lower temperature sQGP. This possibility should be carefully considered in planning future experiments at CERN.

It seems out of question that a new hadronic phase is showing up at RHIC, with quite surprising features, however. Collisions of the initial partons seem to take place from a very peculiar configuration, the Color Glass Condensate~\cite{mclerr}. A very dense, fluid phase results after thermalization:  a strongly interacting Quark Gluon Plasma? In turn, the sQGP raises the issue of which excitations may populate it. 

New phenomena call  for new probes. At RHIC, jet tomography and collective motion have revealed new aspects. Beauty quarkonia could be also very useful as well as muon pairs, to probe hadron resonances surviving in the sQGP.

What about the LHC? 

The structure of the initial partonic state can be investigated by the study of hard jets, or heavy particle production (what about top pairs?) or the Higgs boson (recalling the saying that yesterday's glamour is today's signal... and tomorrow's background). Thermalization of the mid rapidity fireball can be a delicate issue at the large LHC energies: does an asymptotically free fireball have time to thermalize?. 

{\sl Acknowledgments}.

\noindent I would like to thank F. Becattini and A. D. Polosa for interesting discussions on the subjects presented here and for substantial help in the preparation of my talk.


\begin{thebibliography}{9}
\bibitem{WEIN72} S. Weinberg, {\it Gravitation and Cosmology}, John Wiley \& Sons, Inc. New York, London, Sidney, Toronto, 1972.
\bibitem{Green} M.~B.~Green,~J.~Schwarz and E.~Witten, {\it Superstring Theory}, Vol.~1, Cambridge University Press, 1987.
\bibitem{latKar} F. Karsch, {\it Lattice QCD at High Temperature and Density}, hep-lat/0106019.
\bibitem{Bj} J.~D.~Bjorken, Phys. Rev. {\bf D27}, 140 (1983).
\bibitem{NA49_05} NA49 Collaboration, PRL {\bf 94}, 052301 (2005).
\bibitem{brambilla} N. Brambilla {\it et al}., {\it Heavy Quarkonium Physics},~FERMILAB-FN-0779,~CERN-2005-005, Dec 2004. Published as CERN Yellow Report, CERN-2005-005, Geneva: CERN, 2005. -487; hep-ph/0412158.
\bibitem{ellfloSPS} NA49 and NA45(CERES) data, reprted at this Conference.
\bibitem{MPPR_NuclPhys} L. Maiani, F. Piccinini, A. Polosa, V. Riquer, Nucl.Phys {\bf A 748} (2005)209.
\bibitem{Bec}  F.~Becattini, Nucl. Phys. {\bf A702}:~336,~2002.
\bibitem{NA50Pb} M.C. Abreu {\it et al}., Phys. Lett. B450, 456 (1999); M.C. Abreu et al., Phys. Lett. B477, 28 (2000). For the latest analysis see: http://na50.web.cern.ch/NA50/.
\bibitem{NA60In}NA60 Collaboration, presented at $EPS\;05$, Lisbon, and this Conference.
\bibitem{Satzetal}  S. Digal, S. Fortunato, H. Satz, Eur.~Phys.~J. {\bf C32} (2004) 547.
\bibitem{screening} T.~Matsui and H.~Satz Phys. Lett. {\bf B178}, 416 (1986); for a review, see e.g. R. Vogt, Phys. Rep. {\bf 310}, 197 (1999).
\bibitem{ioffe} see e.g. B.L. Ioffe, {\it 10th International Workshop on High-Energy Spin Physics} (SPIN 03), Dubna, Russia, 16-20 Sep 2003, Phys.~Part.~Nucl.{\bf 35} S98-S101,2004 and hep-ph/0310343.
\bibitem{BraunMunz} P.~Braun-Munzinger, {\it Hadron production in ultra-relativistic nuclear
collisions and the QCD phase boundary}, 5th International Conference on Physics and Astrophysics of Quark Gluon Plasma, Salt Lake City, Kolkata, India, 8-12 Feb 2005, nucl-ex/0508024.
\bibitem{MaiCatania} L. Maiani: Concluding Remarks, ALICE-ITALIA, Catania, 12 Jan. 2005.
\bibitem{Bec_al_05} F. Becattini, L. Maiani, F. Piccinini, A. Polosa, V. Riquer, Phys. Lett. B, to appear, and hep-ph/0508188.
\bibitem{shurLNF} E. Shuryak, Seminar given at Laboratori Nazionali di Frascati, Frascati, Italy, Spring 2005, and talk at this Conference, hep-ph/0510123 v2. 
\bibitem{DamjaQM05} S. Damjanovic, QM2005, 4-9 August, Budapest.
\bibitem{RW} R. Rapp and J. Wambach, {\it Chiral Symmetry Restoration and Dileptons in Relativistic Heavy ion Collisions.}, Adv.~Nucl.~Phys. 25:1,2000 and hep-ph/9909229.
\bibitem{BR} G.~E. Brown and M. Rho, Phys. Rev. Lett. 66 (1991) 2720.
\bibitem{mclerr} L.~D. McLerran, R.~Venugopalan, Phys.Rev.~{\bf D49}:2233-2241,1994; hep-ph/9309289.
\end{thebibliography}
\end{document}